\documentclass[%
reprint,
superscriptaddress,
amsmath,amssymb,
prl,
]{revtex4-2}

\usepackage{ulem}%
\usepackage{graphicx}
\usepackage{dcolumn}
\usepackage{bm}
\usepackage{hyperref}

\newcommand{\kB}{k_\mathrm{B}}
\newcommand{\gT}{\gamma_\mathrm{T}}
\newcommand{\gS}{\gamma_\sigma}
\newcommand{\vD}{v_\mathrm{D}}
\newcommand{\pls}{{\footnotesize ($+$)}}
\newcommand{\mns}{{\footnotesize ($-$)}}

\begin{document}
	
	\preprint{APS/123-QED}
	
	\title{Active responsive colloids driven by intrinsic dichotomous noise}
    
	
	\author{Nils G\"oth}
	 \affiliation{Applied Theoretical Physics--Computational Physics, Physikalisches Institut, Albert-Ludwigs-Universit\"at Freiburg, D-79104 Freiburg, Germany}
    \author{Upayan Baul}
     \affiliation{Applied Theoretical Physics--Computational Physics, Physikalisches Institut, Albert-Ludwigs-Universit\"at Freiburg, D-79104 Freiburg, Germany}
	\author{Joachim Dzubiella}
	 \email[Corresponding author: ]{joachim.dzubiella@physik.uni-freiburg.de}
	 \affiliation{Applied Theoretical Physics--Computational Physics, Physikalisches Institut, Albert-Ludwigs-Universit\"at Freiburg, D-79104 Freiburg, Germany}
     \affiliation{Cluster of Excellence livMatS @ FIT--Freiburg Center for Interactive Materials and Bioinspired Technologies, Albert-Ludwigs-Universit\"at Freiburg, D-79110 Freiburg, Germany}
	
	\date{\today}
	
\begin{abstract}
	We study the influence of intrinsic noise on the structure and dynamics of responsive colloids (RCs) which actively change their size and mutual interactions. The colloidal size is explicitly resolved in our RC model as an internal degree of freedom (DOF) in addition to the particle translation. A Hertzian pair potential between the RCs leads to repulsion and shrinking of the particles, resulting in an explicit responsiveness of the system to self-crowding. To render the colloids active, their size is internally driven by a dichotomous noise, randomly switching ('breathing') between growing and shrinking states with a predefined rate, as motivated by  recent experiments on synthetic active colloids. The polydispersity of this dichotomous active responsive colloid (D-ARC) model can be tuned by the parameters of the noise. Utilizing stochastic computer simulations, we study crowding effects on the spatial distributions, relaxation times, and self-diffusion of dense suspensions of the D-ARCs. We find a substantial influence of the 'built-in' intrinsic noise on the system's behavior, in particular, transitions from unimodal to bimodal size distributions for an increasing colloid density as well as intrinsic noise-modified diffusive translational dynamics. We conclude that controlling the noise of internal DOFs of a macromolecule or cell is a powerful tool for active colloidal materials to enable autonomous changes in the system's collective structure and dynamics towards the adaption of macroscopic properties to external perturbations.
\end{abstract}
	
\maketitle
	

\section{Introduction}
For the development of soft functional materials, scientists are typically inspired by proven concepts in nature, as expressed by bacteria, cells, and microswimmers \cite{Buddingh2017, Huang2016}. Unlike inert matter, such as wool or plastic, they all have a power source, like ATP or glucose, which is consumed. And they all have something engine-like built-in which enables an internally controlled change of their appearance or movement. Consequently, their overall behavior and motion is not purely a result of external forces but partially driven by internal mechanisms. Thus, it is not a surprise that active particles have become a large field of interest. The treated topics reach from motile Brownian particles \cite{Farage2015, Romanczuk2012, Bechinger2016} over particles with autonomously oscillating size \cite{Masuda2018, Wessling2021, Tjhung2017} to active (non-motile) hydrogels with pH-feedback \cite{Maity2021}. 

To construct living-like materials, internal activity has to be linked with a large responsiveness to stimuli in the environment~\cite{Walther2020}. For example, bacteria use quorum sensing, emitting and detecting small molecules to evaluate local density, to adapt their internal (genetic, size, speed, etc.) behavior according to their population~\cite{Bassler2001, Wang2015}. Inspired from nature, the approved strategies of responsiveness have been transferred to create artificial materials with controlled response. Widely-investigated are, for example, thermosensitive poly(N-isopropylacrylamide) (PNIPAM) microgels which possess a volume phase transition (switching between collapsed and swollen states) upon small changes in temperature~\cite{Lu2011}, and other stimuli, such as ionic strength \cite{Plamper2017}, pH \cite{Wessling2021, Heckel2021, Cruz-Silva2007} or light \cite{Pasparakis2011}. Potential applications are, e.g., target-oriented drug-delivery where colloids can carry and eject a drug at the diseased tissue \cite{Abulateefeh2011, Minko2010}, or stimuli-responsive, switchable catalysis \cite{Roa2018,Kanduc2020}. 

Responsiveness, switching, and internal activity lead to nonequilibrium distributions and fluctuations on the particle scale, eventually coupled to their collective structure and function. For instance, some bacteria switch randomly between a normal and a persistence state and are therefore resistant to antibiotic treatment due to a small amount of bacteria in the persistence state \cite{Balaban2019, Dubnau2006}. In synthetic systems, block copolymer vesicles show a controlled contraction and expansion behavior upon applying environmental triggers~\cite{Yan2013, Lin2017}. Thereby, a stochastic 'breathing' akin dichotomous noise behavior may be realized. But also autonomously self-pulsating colloids can exhibit irregularities and dichotomous noise-like effects in their oscillations \cite{Narita2013}. In biology, control of internal processes by intrinsic noise gives cells the opportunity to engender heterogeneity in a colony and thereby gain robustness or efficiency \cite{Elowitz2002}. However, these are only few of many examples of stochastic or pseudo-stochastic effects in biology and chemistry \cite{Dittrich2014, Bressloff2017}.

Theoretically, fluctuations and noise are crucial to be well defined in a model, especially in equilibrium the fluctuation-dissipation theorem must hold \cite{BarratHansenBook}. Fluctuations and noise are used to coarse-grain one or multiple microscopic effects which are computationally too expensive to simulate in detail. The Gaussian noise for the translational coordinates typically coarse-grains the interaction with the surrounding solvent ('bath') with a random walk. The fluctuations of active internal degrees of freedom (DOFs) may be far from being Gaussian (white), and we can speak of nonequilibrium, colored noise effects \cite{Caprini2022}. A simple option to violate fluctuation-dissipation and include a kind of activity is to couple different DOFs to different thermostats \cite{Grosberg2015, Dotsenko2013, Netz2020, Polina, Smrek2020}, where one could picture the internal 'engine' as an additional heat bath that can cool or heat the internal DOF. Another possibility is that the particles switch randomly between different kind of states. This can be an active switching between different sizes \cite{Bley2021_switching1, Bley2021_switching2}, between attractive and repulsive particle-particle interactions \cite{Alston2022}, or between different motility states \cite{Sabri2020}. 

In this work, we study the effects of intrinsic dichotomous noise in a model of responsive colloids (RCs) on colloidal structure and dynamics. We recently developed an RC model which resolves besides the translational DOFs an additional internal DOF (or 'property') ~\cite{Lin2020}, for example, representing the colloid's size \cite{Baul2021, bimodal}. The property is assumed to be governed by stochastic dynamics enabling temporal changes and responses to neighboring colloids and other environmental stimuli. The idea of an additional DOF has been used with increasing frequency in the recent years to model complex colloids \cite{Kapteijns2019, Berthier2019, Ciarella2021, Scotti2019, Urich2016} or proteins \cite{Stegen2015}. Other than in previous works \cite{Baul2021, Polina}, we do not drive the internal DOF with a Gaussian noise but with a dichotomous noise. This means that each colloid switches randomly between a growing and a shrinking state. In contrast to known, more coarse-grained and phenomenological models of active switching colloids \cite{Bley2021_switching1, Alston2022} this model is based on a microscopic Hamiltonian and includes also the continuous transition between the two states. The dichotomous noise is one of the simplest switching noises (relevant for the experimental examples above) and has the advantage that it is a very well studied noise \cite{Kitahara1979, VanDenBroeck1983, Morita1990, Masoliver1992, Romanczuk2012, Kim2006, Medeiros2021, Sancho1984}. In particular, it has a known analytical solution for the harmonic potential \cite{Sancho1984}. Using this dichotomous model in stochastic, overdamped simulations, we study intrinsic noise effects on the structure and dynamics of dense colloidal dispersions in the steady state. For an enhanced physical interpretation of the results, we also insert a recapitulation of the existing single-particle solution \cite{Sancho1984} and present a modified perturbation theory for RCs~\cite{bimodal}.

\section{Model and Methods}
\subsection{Dichotomous ARC model (D-ARC)}

To model suspensions of active responsive colloids (ARCs) we utilize our previously introduced RC model~\cite{Baul2021} as basis and modify it with respect to the type of noise for the internal DOF. Consider $N$ colloids with translational temperature $T$, where each colloid (particle) $i$ has a center-of-mass position in 3D space $\boldsymbol{x}_i$. Additionally, each colloid has an associated property $\sigma_i$, in our case study representing a sphere's diameter. This leaves us with in total $4N$ DOFs in the system (three translational and one internal DOF per particle).

For our free energy we consider a single-particle term $U(\sigma)$ and a pair potential term $\phi$. Together we obtain the Hamiltonian 
\begin{equation}
H = \frac{1}{2}\sum_i \sum_{j\neq i} \phi(r_{ij}, \sigma_i, \sigma_j) + \sum_i U(\sigma_i)
\label{eq.hamiltonian}
\end{equation}
where $r_{ij} = \left|\boldsymbol{x}_i - \boldsymbol{x}_j\right|$ is the pair distance between two particles $i$ and $j$.

An appropriate pair potential for soft repulsive and elastic colloids, such as hydrogel particles, is the Hertzian potential \cite{Paloli2013, Rovigatti2019}. The Hertzian interaction potential of two particles is
\begin{equation}
  \phi_{ij} = \phi(r_{ij}; \sigma_i, \sigma_j) = \epsilon
  \left(1 - \frac{r_{ij}}{\tilde{\sigma}} \right)^{5/2}
  \Theta \left(1 - \frac{r_{ij}}{\tilde{\sigma}} \right).
\end{equation}
with the average diameter of both particles $\tilde{\sigma} = (\sigma_i + \sigma_j)/2$ and $\Theta(..)$ denoting the Heaviside-step function. The potential strength is set to $\epsilon = 500~\kB T$ which is found for typical thermosensitive colloids in experiments \cite{Paloli2013}. The pair potential is purely repulsive and cut at $\tilde \sigma$. Thus, not overlapping particles do not interact and the larger the overlap the higher the potential energy. A more detailed discussion about the Hertzian potential can be found elsewhere \cite{Paloli2013, Baul2021}.

In addition to the pair potential, RCs feature the internal parent energy landscape, $U(\sigma)$, for the evolution of the size property $\sigma$. For simplicity, we choose $U$ to be a harmonic potential of the form $U(\sigma) = \frac{1}{2\beta\delta^2}(\sigma-\sigma_0)^2$ with a potential width of $\delta = 0.2\sigma_0$ and $\beta = 1/\kB T$ being the inverse thermal energy. This defines a preferred particle size of $\sigma_0$ and avoids very small and very large particle sizes due to its confinement. A Gaussian parent implies a simple linear elastic response of the size, while for hydrogels more accurate nonlinear responses can also be straightforwardly employed \cite{Scotti2019}.

The translational force $\boldsymbol{F}_i^\mathrm{T}$ acting on a particle $i$ is given by the gradient of the Hamiltonian in Eq.~(\ref{eq.hamiltonian}). We transfer this procedure to the property coordinate $\sigma$ which leads to the forces
\begin{eqnarray}
	\boldsymbol{F}_i^\mathrm{T} &= -\nabla_i H &= -\sum_{j\neq i} \nabla_i \phi_{ij},
	\label{eq.model_force_x}\\
	F_i^\sigma &= -\partial_{\sigma,i}H &= -\partial_{\sigma, i} U - \sum_{j\neq i} \partial_{\sigma, i} \phi_{ij},
	\label{eq.model_force_sig}
\end{eqnarray}
where $\nabla_i$ and $\partial_{\sigma, i}$ are the derivatives with respect to the translational coordinates and the property coordinate, respectively. Since the pair potential $\phi_{ij}$ also affects the property, an overlap of two particles leads not only to a repulsion but also to a shrinking of both particles~\cite{Baul2021}. Thus, the particles respond internally to their local environment.

Regarding the equations of motions, as in \cite{Baul2021} we neglect the inertia term for the viscous and stochastic motion of all DOF of the RCs, leading to overdamped Langevin-like dynamics. To solve the stochastic differential equations we chose the basic Euler algorithm which is computationally slower but simpler compared to more sophisticated methods like the Runge-Kutta or Milstein scheme \cite{KloedenPlatenBook}. The discretized version for the position and property reads (It\^{o} convention \cite{KloedenPlatenBook})
\begin{eqnarray}
\label{eq.eom_x}
\boldsymbol{x}_i(t + \Delta t) &= \boldsymbol{x}_i(t) &+ \frac{\Delta t}{\gamma_\mathrm{T}}\boldsymbol{F}_i^\mathrm{T}(t) + \sqrt{\frac{2\kB T\Delta t}{\gamma_\mathrm{T}}} \boldsymbol{\xi}_i^\mathrm{T}(t),\\
\sigma_i(t + \Delta t) &= \sigma_i(t) &+ \frac{\Delta t}{\gamma_\sigma} F_i^{\sigma}(t) + \Delta t \mathcal{I}(t).
\label{eq.eom_s}
\end{eqnarray}
Here is $\Delta t$ the simulation timestep, $\gamma_\mathrm{T}$ and $\gamma_\sigma$ the translational and property friction coefficient and $\bm{F}_i^{\mathrm{T}}$, $F_i^\sigma$ are the forces from Eqs.~(\ref{eq.model_force_x}), (\ref{eq.model_force_sig}). In this overdamped dynamics, the random translational movement of a particle is described by Stokes friction $\gamma_\mathrm{T} \propto \sigma$ and the conventional Gaussian (white) noise with $\left< \boldsymbol{\xi}_i(t) \right> = 0$ and $\left< \xi_{i, k}^\mathrm{T}(t) \xi_{j, \ell}^\mathrm{T}(t') \right> = \delta_{ij} \delta_{k\ell} \delta(t-t')$. Here is $k, \ell \in \left\{x, y, z\right\}$ and $\delta$ being the Kronecker delta and the Dirac $\delta$-function, respectively. Since the size $\sigma$ is a dynamically changing variable, we assume that the {\it ad hoc} friction of a particle is calculated by $\gamma_\mathrm{T}(\sigma) = \gamma_\mathrm{T}^0 \sigma/\sigma_0$ with $\gamma_\mathrm{T}^0 = \gamma_\mathrm{T}(\sigma_0) = 1\frac{\tau}{\beta\sigma_0^2}$. This also sets the single-particle diffusion for fixed size $\sigma_0$ through the Stokes-Einstein equation $D^0_\mathrm{T} = 1/(\beta \gamma_\mathrm{T}^0)$. We define the unit of length $\sigma_0 \equiv 1$, the unit of time $\tau \equiv 1$ and the unit of energy $1/\beta \equiv 1$. Moreover, we fix the property friction coefficient (setting the time scale of the size relaxation) to $\gamma_\sigma = 1000\gT^0$ which is therefore much higher than for the translation and assures that diffusion happens on a faster timescale than size changes~\cite{Baul2021}.

As motivated in the Introduction, we activate our particles internally by using a dichotomous noise $\mathcal{I}(t)$ for the random behavior of the size, $\sigma$. The equations (\ref{eq.eom_x}) and (\ref{eq.eom_s}) define therefore our dichotomous ARC model (D-ARC). The features of the dichotomous noise $\mathcal{I}(t)$ and its consequences on single particle behavior will be discussed in detail in the subsequent section. If $\mathcal{I}(t)$ were replaced by a Gaussian noise with temperature $T$, the system would be identical to the equilibrium RC system studied previously with Brownian dynamics for all DOFs~\cite{Baul2021}. We note that the active case of an internal Gaussian noise with a temperature different than the translational temperature (i.e., different DOFs coupled to two different temperature baths) was studied by Gaindrik {\it et al.} \cite{Polina}, exhibiting rich nonequilibrium effects.

\subsection{Dichotomous noise}
\label{sec.dichotomous_noise}

\begin{figure}[b]
	\centering
	\includegraphics[width=8.6cm]{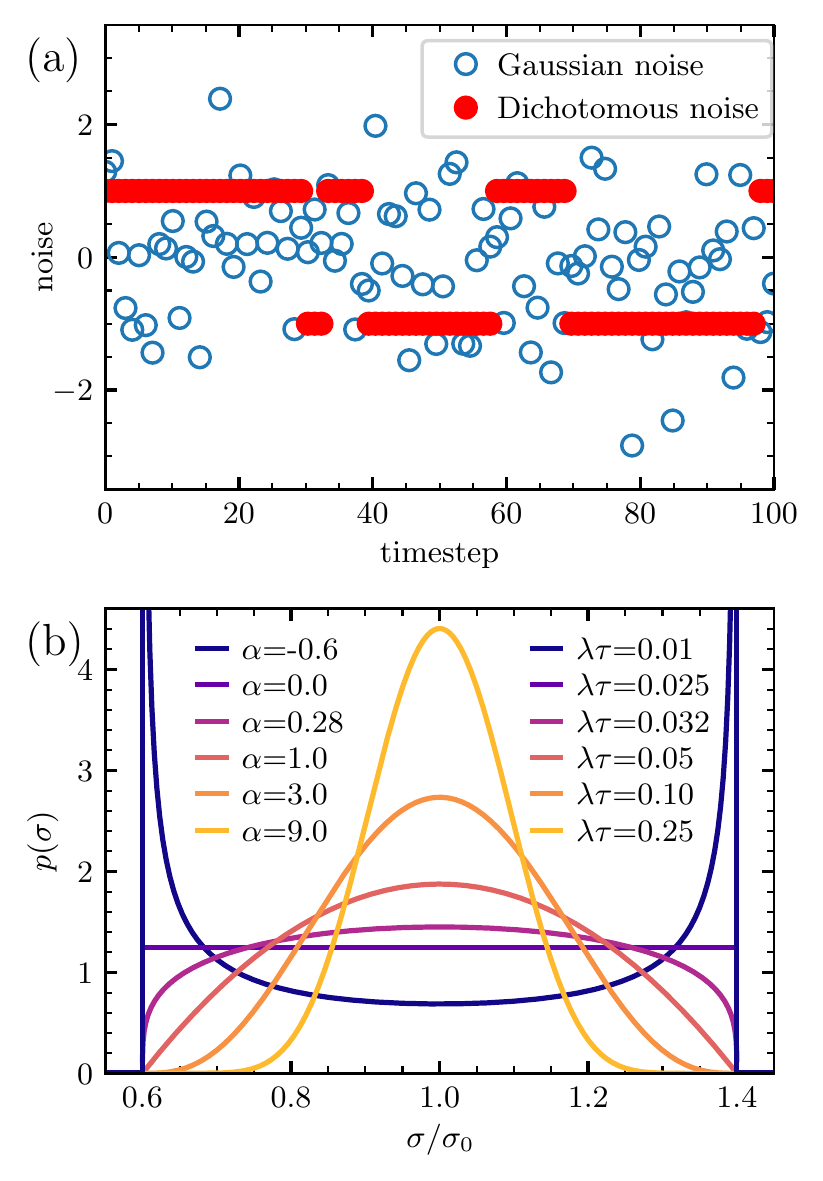}
	\caption{Dichotomous noise. (a) Comparison of Gaussian noise (blue circles) and dichotomous noise (red dots). For both the noise values for a series of 100 timesteps are shown. The variances are normalized to 1 and the dichotomous process has a switching rate of $\lambda =$ 0.05/timestep and therefore at each timestep a probability of $e^{-0.05} \approx 0.95$ to remain in the current state. (b) Parent property distribution $p(\sigma)$ for a constant dichotomous velocity and different exponents $\alpha$. The denoted switching rates are related to $\alpha$ via Eq.~(\ref{eq.dm_propDist}) and correspond to the parameters used in the simulation.}
	\label{fig.dm}
\end{figure}

In contrast to Gaussian noise, dichotomous noise is discrete and can only take two values $\mathcal{I} = \pm \vD$ (see Fig.~\ref{fig.dm}(a)). If a particle receives the value $+\vD$, we call it to be in the \pls-state and for $-\vD$ it is in the \mns-state. Due to the fact that $\vD$ has units of a velocity, we call it {\it dichotomous velocity}. It is the velocity with which a particle in absence of other forces grows or shrinks. Another difference to the Gaussian noise is that it is not $\delta$-correlated in time. More precisely, the dichotomous noise is defined via \cite{Morita1990}
\begin{eqnarray}
    \left<\mathcal{I}(t)\right> &=& 0,\\
    \left<\mathcal{I}(t)\mathcal{I}(t')\right> &=&
    \vD^2 e^{-2\lambda|t-t'|},
    \label{eq.dm_acf}
\end{eqnarray}
where $\lambda$ denotes the switching rate between the two states. The switches happen purely randomly in time. The probability that a particle is  still in the same state after a time $t$ (no switch) is given by the Poisson process $P(t) = \exp[-\lambda t]$. Since only the last noise value has an influence on the next one, it is still a Markov process, why it is also called dichotomous Markov process. It shall be noted that the notation in literature is not consistent and the prefactor of $2$ in the exponent of Eq.~(\ref{eq.dm_acf}) is sometimes omitted.

\subsection{Parent distribution under dichotomous noise}

In the following we discuss analytical solutions of the isolated, single-particle case, or low density limit (LDL), of our model, where particle-particle interactions are neglected. However, the property distribution of the DOF, $p(\sigma$), in a harmonic potential under dichotomous noise is non-trivial, since it is not simply given by the Boltzmann distribution $p\propto \exp(-\beta U)$. For an isolated colloid, the property equation of motion from Eq.~(\ref{eq.eom_s}) simplifies to
\begin{equation}
    \dot{\sigma} =
    f + g\mathcal{I} \quad \mathrm{with} \quad f(\sigma) = -\frac{1}{\gamma_\sigma}\partial_\sigma U(\sigma), \quad g(\sigma)=1.
    \label{eq.dm_singleparticle_eom}
\end{equation}
Kitahara {\it et al.} found in an impressive work \cite{Kitahara1979} a stationary solution of Eq.~(\ref{eq.dm_singleparticle_eom}) for natural boundaries and general functions $f(\sigma)$ and $g(\sigma)$:
\begin{eqnarray}
    p(\sigma) &=& p_0 \frac{g(\sigma)}{v_\text{D}^2 g^2(\sigma) - f^2(\sigma)} \nonumber\\
    & &\times \exp \left[ 2\lambda \int^{\sigma} \mathrm{d}\sigma'
    \frac{f(\sigma')}{v_\text{D}^2 g^2(\sigma') - f^2(\sigma')} \right]
    \label{eq.dm_singleparticle_Nfg}
\end{eqnarray}
with a normalization factor $p_0$. By inserting our $f(\sigma)$ and $g(\sigma)$ from Eq.~(\ref{eq.dm_singleparticle_eom}) into Eq.~(\ref{eq.dm_singleparticle_Nfg}), we obtain for the stationary probability distribution \cite{Sancho1984}
\begin{equation}
	p(\sigma) = \left\{\begin{array}{ll} p_0 \left(1 - \left(\dfrac{\sigma-\sigma_0}{\Delta}\right)^2 \right)^\alpha &, \sigma\in\left(\sigma_0-\Delta, \sigma_0+\Delta\right) \\
	0 &, \mathrm{otherwise}\end{array}\right. \nonumber
    \label{eq.dm_propDist}
\end{equation}
\begin{equation}
    \mathrm{with} \quad \Delta = v_\text{D} \beta\gamma_\sigma \delta^2, \quad
    \alpha = \lambda \beta\gamma_\sigma \delta^2-1. \quad 
    \label{eq.dm_propDist_Delta_alpha}
\end{equation}
and $p_0 = \Gamma(\alpha+3/2)/(\sqrt{\pi}~\Gamma(\alpha+1))$ being a prefactor normalizing the distribution to 1 which contains the gamma function $\Gamma$. The parameter $\Delta$ defines the width of the distribution which is only non-zero in the interval $\sigma \in \left(\sigma_0 - \Delta, \sigma_0 + \Delta\right)$. At both boundaries the force due to the single-particle potential and the dichotomous force cancel. Consequently, the particles have a minimum and a maximum size which is identical to the distribution's boundaries. The dichotomous velocity $\vD$ is proportional to the width $\Delta$. The second parameter describing the property distribution is the exponent $\alpha$ which depends on the switching rate $\lambda$ of the noise.

The property distribution for different values of $\alpha$ can be seen in Fig.~\ref{fig.dm}(b). In the limit of no switching ($\lambda \rightarrow 0,$ $\alpha \rightarrow -1$) the distribution consists of two Dirac $\delta$-functions, one at each boundary, which is a simple bidisperse system. For low switching rates ($\alpha < 0$), the DOF develops a probability distribution with two finite-width peaks and a lower probability in between, which is a bimodal distribution. The transition state to unimodality (one peak) is a uniform distribution ($\alpha = 0$). Further rise of the switching rate ($\alpha > 0$) results in an increasingly sharp peak in the center. In the limit of infinitely fast switching ($\lambda \rightarrow \infty$, $\alpha \rightarrow \infty$) we obtain one Dirac $\delta$-function at $\sigma_0$, characterizing a monodisperse system. Gaussian white noise is obtained in the limit $\lambda \rightarrow \infty$, $v_\mathrm{D} \rightarrow \infty$ with $\lambda/ v_\mathrm{D}^2 = \beta \gamma_\sigma /2$.

Already the single-particle solution is interesting from a physical point of view of (bio)chemical matter. It enables colloids or bacteria to control their own size distribution by {\it internally} tuning their switching rate (and swelling velocity) which is important for function~\cite{Cesar2017}. We focus in this work on the influence of the switching rate $\lambda$ which is an important parameter for biological systems, like bacteria, to regulate switching between different phenotypes \cite{Dubnau2006}. However, note that the control of the parameter $\vD$ would also be quite powerful since it is in the LDL directly proportional to the distribution width.

Beyond the LDL, at higher densities, particles interact and the distribution $p(\sigma)$ will be modified. We thus distinguish in our work between the property distribution in the LDL $p(\sigma)$ (the 'parent') and the {\it emergent} property distribution for non-vanishing particle densities, $N(\sigma)$.

\subsection{Simulation details}
\begin{figure}
	\centering
	\includegraphics[width=8.6cm]{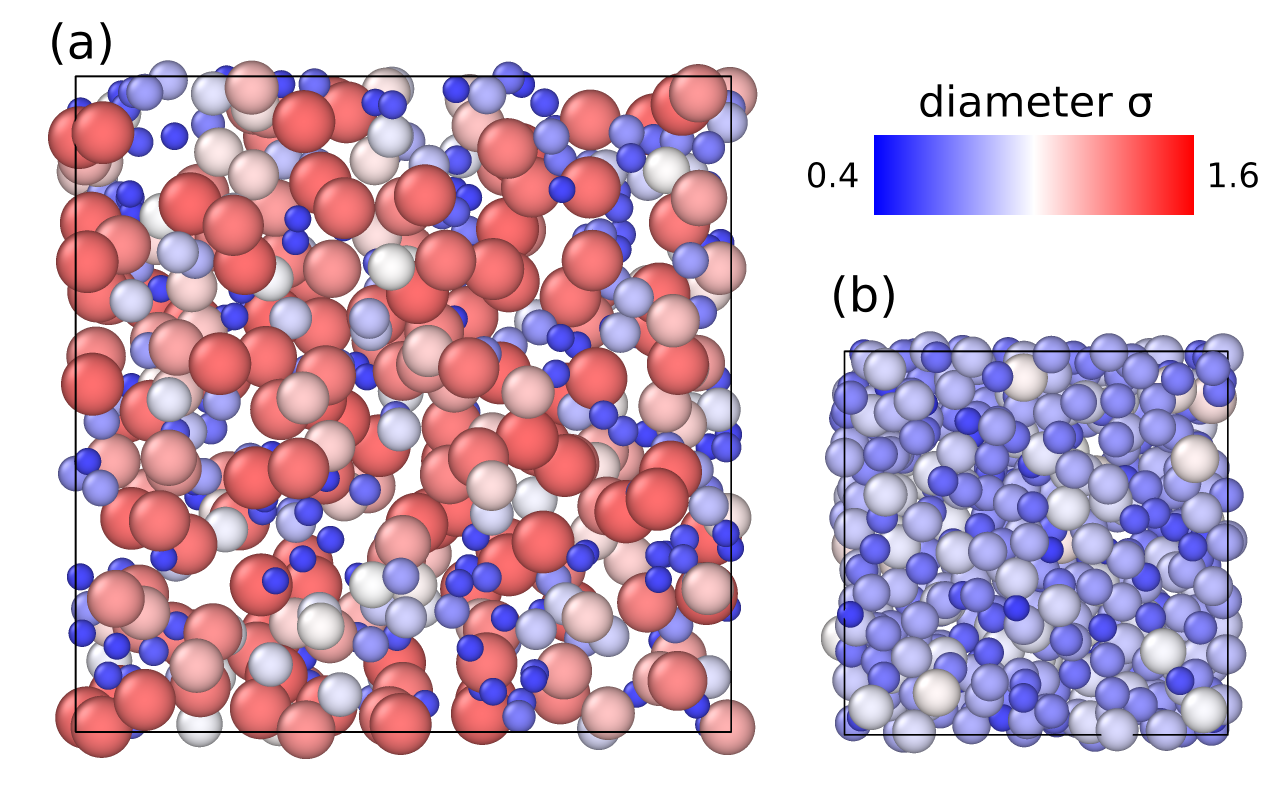}
	\caption[Simulation snapshots]{Snapshots of the simulation. (a) Low particle density ($\rho\sigma_0^3 = 0.19$) and low switching rate ($\lambda\tau = 0.01$). (b) High particle density ($\rho\sigma_0^3 = 0.95$) and high switching rate ($\lambda\tau = 0.1$). The box with periodic boundary conditions is depicted by thin black lines. The particle's color visualizes its size; red indicates a large particle, blue a small one and a white particle has a size of $\sigma = \sigma_0$. The two boxes are scaled to the same reference length; the left box has a side length of $L \approx 14\sigma_0$ and the right box of $L \approx 8\sigma_0$, respectively. Snapshots are made with OVITO \cite{OVITO}.}
	\label{fig.snapshot}
\end{figure}

The computer simulations leading to the numerical results contain $N=512$ particles in a cubic box with periodic boundary conditions. To obtain a certain particle number density $\rho$, the length of the box is fixed to $L = \sqrt[3]{N/\rho}$. Simulation snapshots for two different densities are shown in Fig.~\ref{fig.snapshot}. Initially all particles are distributed on a simple cubic lattice filling in the whole box. Simultaneously, each particle receives a random state (\pls~or \mns) and a size $\sigma_i$ pulled from a Gaussian distribution centered at $\sigma_0$ and with variance $\delta^2$.

For the time evolution a simple Euler algorithm is used (cf. Eqs.~(\ref{eq.eom_x}), (\ref{eq.eom_s})). The duration of one timestep is $\Delta t = 10^{-4}\tau$. Each simulation contains $3\times10^6$ (or $5\times10^6$) equilibration steps, to ensure that the system is in the steady state, and subsequent $10^7$ production steps, which are used to collect data. The configuration is written out every 1000th timestep. For each analyzed parameter set five independent simulations were performed; all results are averages over these five data sets. With $\gS = 1000\gT$ and a dichotomous velocity of $\vD = 0.01\sigma_0/\tau$ the property motion is much slower than the translational motion. This choice leads to internal relaxations much slower than translation and is motivated by small, chemically-stimulated microgels for which translational diffusion happens much faster than size changes due to the complex internal processes \cite{Xing2011}. It is also known in general that internal degrees of freedom can slow down internal relaxation beyond idealized hydrodynamic conformational behavior, for example, in the slowing down of polymer folding rates due to internal friction processes \cite{Cheng2013}. All model and simulation parameters are summarized in Tab.~\ref{tab.mod_sim_parameters}.

To implement the dichotomous noise we make use of method 2 presented in \cite{Kim2006}. Each particle gets initially a random remaining time $t_\mathrm{r} = -\ln(u)/\lambda$, where $u$ is drawn from a uniform distribution between 0 and 1. After the remaining time is over, the particle switches its state and a new remaining time, drawn in the same way, is assigned to the particle. This procedure leads to the same noise as deciding at each timestep whether a particle switches its state or not, but saves some computational time.

The used random number generator for a uniform distribution between 0 and 1 is the Mersenne Twister MT19937, followed by the Marsaglia polar method \cite{Marsaglia1964} to obtain random numbers drawn from a Gaussian distribution.

\begin{table}[]
	\centering
	\renewcommand*{\arraystretch}{1.2}
	\begin{tabular}{>{\centering}p{1.0cm}>{\centering}p{3.0cm}<{\centering}p{4.2cm}}
		\toprule
		Parameter & Value & Description \\ \hline
		$\sigma_0$ & $1$ & unit length and mean size of an isolated particle \\
		$\tau$ & 1 & unit time defined by translational diffusion $D_\mathrm{T}^0 = 1 ~\sigma_0^2/\tau$ \\
		$\beta$ & 1 & inverse unit energy ($\kB T = 1/\beta$) \\ \hline
		$\delta$ & $0.2\sigma_0$ & width of $U(\sigma)$ \\
		$\gamma_\mathrm{T}^0$ & $1 ~{\tau}/{(\beta\sigma_0^2)}$ & translational friction coefficient of a particle with size $\sigma_0$\\
		$\gS$ & $10^3 ~{\tau}/{(\beta\sigma_0^2)}$ & property friction coefficient\\
		$\vD$ & $0.01 ~{\sigma_0}/{\tau}$ & dichotomous velocity \\
		$\beta \epsilon$ & $500$ & Hertzian potential strength\\ \hline
		$\lambda\tau$ & $\lbrace 0.01, 0.025,$ $0.032, 0.05,$ $0.1, 0.25 \rbrace$ & (dichotomous) switching rate \\
		$\rho\sigma_0^3$ & $\lbrace 0.019, 0.19,$ $0.57,0.95,$ $1.33 \rbrace$ & number density of particles\\ \hline
		$N$ & $512$ & number of particles\\
		$\Delta t$ & $10^{-4}\tau$ & length of one timestep \\
		-- & $~300\tau-500\tau$ & equilibration time\\
		-- & $1000\tau$ & production time\\
		-- & $10/\tau$ & output frequency\\
		\toprule
	\end{tabular}
	\caption{Parameters in the numerical simulation, their values and their description. The table is divided into four parts: unit sizes, model parameters, varied parameters, and simulation specific parameters.}
	\label{tab.mod_sim_parameters}
\end{table}

\subsection{Radial distribution function and structure factor}
A common measure to analyze the structure of a suspension is the radial distribution function (RDF), $g(r)$. It correlates the pair distances between particles and is defined as~\cite{BarratHansenBook} 
\begin{eqnarray}
	g(r) = \frac{1}{4\pi r^2\rho N(N-1)} \left< \sum_i \sum_{j\neq i} \delta\left( r - r_{ij} \right) \right>
	\label{eq.gofr}
\end{eqnarray}
with $r_{ij} = \left|\boldsymbol{x}_i - \boldsymbol{x}_j\right|$. Since we can split our particles into two groups (\pls- and \mns-state particles), we can also split the RDF into its components, via
\begin{equation}
	g(r) = \frac{1}{4}g_{-}(r) + \frac{1}{4}g_{+}(r) + \frac{1}{2}g_{\pm}(r),
	\label{eq.gofr_split}
\end{equation}
where $g_{-}(r)$ and $g_{+}(r)$ are the RDFs of only the \mns-state and \pls-state particles, respectively. Meanwhile, $g_{\pm}(r)$ includes only distances of particles with different states. We also calculate the structure factor \cite{AllenTildesley2017}
\begin{equation}
S(q) = 1 + 4\pi\rho \int_{0}^{\infty} \mathrm{d}r~ r^2 \frac{\sin(qr)}{qr} \left[g(r)-1\right].
\label{eq.sofq_simple}
\end{equation}
which can be obtained by scattering experiments \cite{BarratHansenBook}.

\subsection{Diffusion coefficient}

For characterizing the translational diffusion in our systems we compute the spatial mean-squared displacement (MSD)
\begin{eqnarray}
	\mathrm{MSD}(t) \equiv \left<\left(\boldsymbol{x}(t) - \boldsymbol{x}(0)\right)^2\right>.
	\label{eq.msd}
\end{eqnarray}
For long times the MSD is proportional to $t$ and we can obtain an effective diffusion coefficient through fitting via $\mathrm{MSD}(t) = 6D_\mathrm{T}^\mathrm{eff}t$. Only the values within the interval $t \in \left[ 10\tau, 250\tau\right]$ are used for the fit to omit the short time diffusion and the MSD values with small statistics in the case for long times.

\subsection{Property auto-correlation function}

A common measure to access the dynamics and relaxation times of a DOF is the auto-correlation function (ACF). We define the normalized ACF of the property as
\begin{equation}
	C_{\sigma\sigma}(t) = \frac{\left<\sigma(t)\sigma(0)\right> - \left<\sigma\right>^2}{\left<\sigma^2\right> - \left<\sigma\right>^2},
	\label{eq.acf_norm}
\end{equation}
which starts at $C_{\sigma\sigma}(t=0)=1$, and converges to 0 in the long-time limit because initial and final value of $\sigma$ are uncorrelated due to the random processes. The ACF for one dichotomous particle in a harmonic potential, which corresponds to our LDL, can be determined analytically~\cite{Sancho1984} and reads
\begin{equation}
	C_{\sigma\sigma}^\mathrm{LDL}(t) = \frac{1}{t_\delta - t_\mathrm{D}} \left[t_\delta e^{-t/t_\delta} - t_\mathrm{D} e^{-t/t_\mathrm{D}}\right]
	\label{eq.acf_LDL}
\end{equation}
with the time constants $t_\delta = \beta\gamma_\sigma\delta^2$ and $t_\mathrm{D} = 1/2\lambda$. We see that the normalized ACF is sum of two independent exponential decays. The potential time $t_\delta$ describes the time scale on which a noise-less particle moves to the center of the single-particle potential. The dichotomous time $t_\mathrm{D}$ characterizes the time scale of the mean lifetime of a state. Both processes are independent of each other. Finally, to quantify the decay's speed we use the correlation time $t_\mathrm{corr}$ which is defined by
\begin{equation}
	C_{\sigma\sigma}(t_\mathrm{corr}) = 1/e.
	\label{eq.acf_corrTime}
\end{equation}
In equilibrium systems the time derivative of the unnormalized auto-correlation function is proportional to the response function (fluctuation-dissipation theorem) \cite{Medeiros2021}. This relation was already successfully used in non-equilibrium systems \cite{Fodor2018} which gives us access to a utilizable response function.

\section{Results}
\label{sec.results}

\begin{figure*}[ht!]
	\centering
	\includegraphics[width=17.9cm]{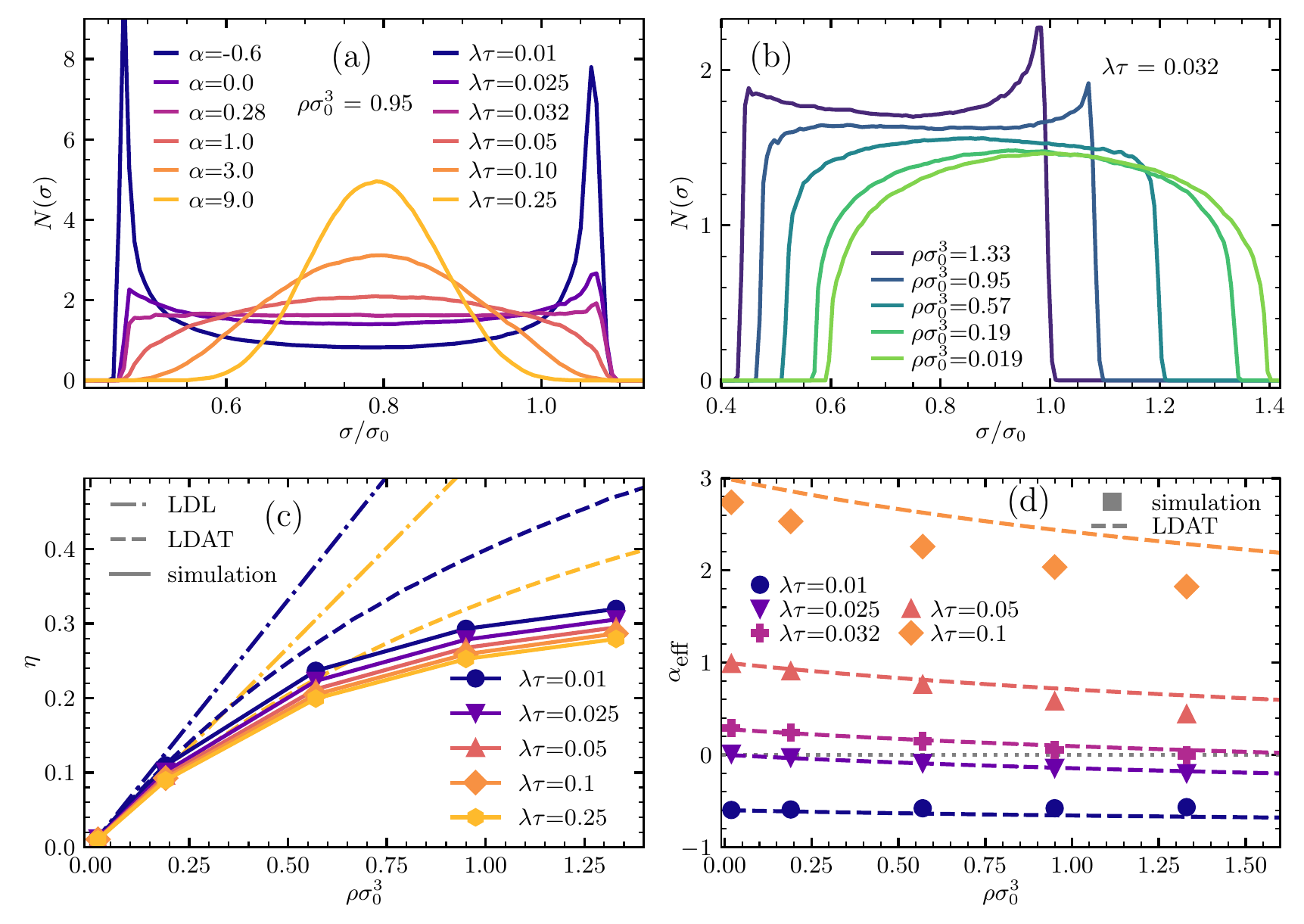}
	\caption{\label{fig.propDist_rho095} (a) Emergent property distribution $N(\sigma)$ for a density of $\rho\sigma_0^3 = 0.95$ and different switching rates. The displayed $\alpha$ denote the exponent in the LDL which can be converted to $\lambda$ using Eq.~(\ref{eq.dm_propDist_Delta_alpha}). The values close to the edges are impaired by the binning algorithm and have to be treated with caution. (b) Property distribution $N(\sigma)$ for a constant switching rate and different particle densities. One can see a transition from a unimodal to bimodal for increasing densities. (c) Packing fraction $\eta = \pi\rho\langle\sigma^3\rangle/6$ vs. number density for different switching rates. The symbols and solid lines depict the simulation results, the dashed lines the LDAT and the dash-dotted line the linear scaled LDL. (d) The symbols depict the effective exponent $\alpha_\mathrm{eff}$ from fitting Eq.~(\ref{eq.dm_propDist}) to the obtained property distribution. The dashed lines show again the prediction from the LDAT.}
\end{figure*}

\subsection{Property distribution}

\begin{figure*}
	\centering
	\includegraphics[width=17.9cm]{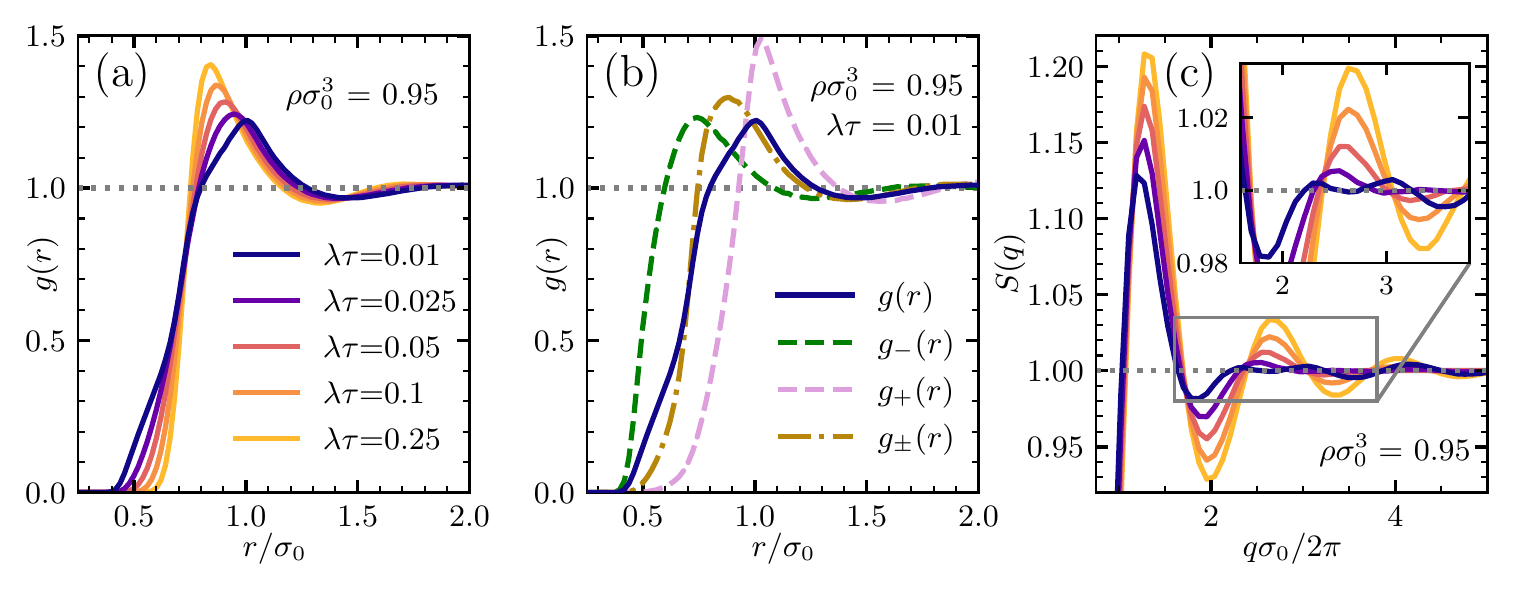}
	\caption{\label{fig.gofr}
		(a) RDF for a number density of $\rho\sigma_0^3 = 0.95$ and different switching rates. (b) The $g(r)$ can split into its components according to Eq.~(\ref{eq.gofr_split}). The dashed blue line shows the total RDF, the green (violet) dashed line shows the $g(r)$ of only the \mns- (\pls-) state particles; $g_{\pm}(r)$ (golden dash-dotted line) is for RDF between particles of different states. (c) Structure factor calculated from the RDFs shown in (a) with the same colors as in (a).}
\end{figure*}

We first examine property distributions for non-dilute particle densities: In Fig.~\ref{fig.propDist_rho095}(a) the emergent property distribution $N(\sigma)$ is shown for $\rho\sigma_0^3 = 0.95$ and different switching rates. Although the system is quite dense, we can see clear similarities to the low-density limit (LDL). The edges of the bimodal distributions are still very sharp because of the comparatively slow movement in the $\sigma$-dimension. Consequently, a particle's $\sigma$-movement can be seen as in a mean potential established by the other particles in addition to the single-particle potential. However, an interesting phenomenon can be extracted from the $\alpha = 0$-distribution in Fig.~\ref{fig.propDist_rho095}(a) which corresponds to the uniform distribution in the LDL (cf. Fig.~\ref{fig.dm}(b)). The increase in density shifts the flat distribution towards bimodality. This can more clearly be seen in Fig.~\ref{fig.propDist_rho095}(b) which shows also the property distribution but only for $\lambda\tau = 0.032$ and different densities in return. Four points shall be stressed from this figure: i) For increasing density the distribution's center shifts to lower values of $\sigma$ (cf. snapshot in Fig.~\ref{fig.snapshot}(b)), while ii) the distribution gets narrower at the same time. Both i) and ii) are also observed in \cite{Baul2021} where a Gaussian noise is used for the property instead of the dichotomous noise. iii) The distribution makes a transition from unimodality to bimodality. While it is for this specific $\lambda$ unimodal for low densities (e.g., $\rho\sigma_0^3 = 0.019$), it is bimodal for higher densities (e.g., $\rho\sigma_0^3 = 1.33$). Thus, it is a density-induced transition. iv) We observe asymmetries for higher densities. For $\rho\sigma_0^3 = 0.95$ one can see that the right edge shows an indication of bimodality while the left edge has a unimodal shape. A look on Fig.~\ref{fig.propDist_rho095}(a) shows that these asymmetries are comparatively small for the covered densities.

The crowding effect itself is well described with the mean packing fraction $\eta = \pi \rho \langle \sigma^3 \rangle /6$ which is the volume fraction filled by the colloids. The packing fraction is shown in Fig.~\ref{fig.propDist_rho095}(c) vs. number density. The simulation results of our RCs show a sub-linear behavior of $\eta(\rho)$ due to the colloid shrinking in crowded environments~\cite{Baul2021}. The spread with respect to the switching rate results from the different property distribution shapes: low switching rates privilege extreme sizes while a small and a large colloid fill more space than two average-sized colloids ($\sigma^3$-dependency). Therefore, $\eta$ increases with decreasing $\lambda\tau$. We complement our result with a perturbation theory for RCs~\cite{Baul2021}. Briefly, we make a low density assumption and calculate the mean force on a particle with size $\sigma$. By a perturbation of the free energy landscape $U(\sigma)$ with non-zero particle densities we obtain our low density approximation theory (LDAT). Details and equations are shown in the Appendix A. Predictions from our LDAT for the packing fraction can exemplary be seen as dashed lines in Fig.~\ref{fig.propDist_rho095}(c). Even though the theory underestimates the crowding effect for high densities, it exhibits the sub-linear behavior which is the crucial point.

As we saw in Fig.~\ref{fig.propDist_rho095}(a), the property distribution's shape for higher densities is still pretty similar to the one in the LDL described by Eq.~(\ref{eq.dm_propDist}). Therefore, we want to fit Eq.~(\ref{eq.dm_propDist}) to the obtained distribution and determine an effective value for $\alpha$, which then describes the distribution's shape and the degree of unimodality/bimodality. The left and right edge of a distribution are identified by taking the first $\sigma_\mathrm{f}$ and last value $\sigma_\mathrm{l}$ of the normalized distribution ($\int \mathrm{d}\sigma N(\sigma) = 1$) crossing $N(\sigma) = 0.001$. With these parameters, the new center $\sigma_1 = (\sigma_\mathrm{f} + \sigma_\mathrm{l})/2$ and width $\Delta_1 = (\sigma_\mathrm{l} - \sigma_\mathrm{f})/2$ of the distribution can be calculated. Eq.~(\ref{eq.dm_propDist}) with given $\sigma_1$ and $\Delta_1$ can now be fitted to the obtained distribution with $\alpha$ as fit parameter. The emerged value is the effective exponent $\alpha_\mathrm{eff}$. Fig.~\ref{fig.propDist_rho095}(d) shows $\alpha_\mathrm{eff}$ as a function of density for different switching rates. We see for all covered switching rates a decrease of $\alpha_\mathrm{eff}$ for rising $\rho$. This illustrates the observed transition from unimodality ($\alpha_\mathrm{eff} > 0$) to bimodality ($\alpha_\mathrm{eff} < 0$).

The basis for the uni-to-bimodal transition lies in the mean potential seen by a particle. In the LDL we have for the property only the quadratic single-particle potential. The additional potential, arising from particle-particle interactions, shifts the total potential to lower $\sigma$ values and narrows it. While the dichotomous velocity remains unmodified, the distance between minimum and maximum $\sigma$ shortens. Hence, more particles gather at the distribution's edges which induces a trend to bimodality. This bimodality is not a consequence of barrier crossing in the free energy landscape as in our previous RC model \cite{bimodal}, but a result of the underlying noise as in \cite{Caprini2022,Fodor2018}. The LDAT (see Appendix A) can now be used to approximate the property distribution for non-vanishing densities. The $\alpha_\mathrm{eff}$ predicted from the LDAT is also shown in Fig.~\ref{fig.propDist_rho095}(c). It describes the trend quite good for low densities and also predicts the density-induced transition. For higher densities, the LDAT underestimates the transition which is a result of the neglect of higher order interaction terms in the theory.

Regarding the physical interpretation of our results, we speculate that the observed density behavior may enable active colloidal suspensions to perform a collective response to changes in density without communication via chemical signaling. This could be interesting in combination with autonomously, self-oscillating particles where the occurrence of the oscillations can highly depend on the density and numbers of neighboring 'coupling' particles~\cite{Wessling2021}. The unimodal-bimodal transition may also engender automatic changes in size diversity depending on the density. This could be an important factor in the development of adaptive materials \cite{Cesar2017} based on synthetic active colloidal dispersions.

\subsection{RDF and structure factor}

We now analyze the spatial particle-particle correlations by inspecting radial distribution functions (RDFs). Fig.~\ref{fig.gofr}(a) shows the RDFs for $\rho\sigma_0^3 = 0.95$ and different switching rates. The observed behavior for high $\lambda\tau$ is as expected for common monodisperse liquids \cite{HansenMcDonaldBook}. However, for $\lambda\tau = 0.01$ we observe a clear deviation, consisting of a substructure of three washed out step functions. The explanation can be found when splitting the RDF into its individual components (cf. Eq.~(\ref{eq.gofr_split})); as done in Fig.~\ref{fig.gofr}(b). If we consider only the \mns-state particles in our system (green dashed line), we spot a peak which is located at $r_\mathrm{p} \approx 0.7\sigma_0$; this is much lower than the peak of the total $g(r)$ ($r_\mathrm{p} \approx 1.0\sigma_0$). The reason is that \mns-state particles are in general much smaller than the average. The same can be said for \pls-state particles (violet dashed line) with the difference that the peak is at further distances. Only $g_{\pm}(r)$ shows an approximately `normal' behavior with a peak in between the others. Since the three peaks have a comparably large distance and the total RDF is the sum of these three ($g_{\pm}(r)$ counts double), $g(r)$ shows a substructure. Thus, the appearance of the substructure is not a result of the particles' dynamics but simply of the bimodal property distribution. Note that a substructure is present for all switching rates, but is in general not visible in the RDF due to too close peaks.

To visualize the substructure we plot the corresponding structure factor (see Eq.~(\ref{eq.sofq_simple})) of Fig.~\ref{fig.gofr}(a) in Fig.~\ref{fig.gofr}(c). For all $\lambda\tau$ we see a first peak at $q\sigma_0 / 2\pi \approx 1.2$ resulting from the presence of the first coordination shell. The second peak shows the substructure: while it is clearly visible for high switching rates, it is suppressed for $\lambda\tau = 0.01$. A similar but weaker effect is also visible for $\lambda\tau = 0.025$, where the second minimum is suppressed. The approximately identical behavior for $q\sigma_0 / 2\pi < 1$ for all $\lambda\tau$ implies a very similar long range order while losing short range structure ($q\sigma_0 / 2\pi > 1$) is recognized for high $\lambda\tau$. This can be explained by the broad property distribution which leads to many different occurring short particle-particle distances (see also Fig.~\ref{fig.gofr}(a)).

We tentatively conclude that physically the intrinsic noise and internal distributions have a major global effect and can serve as control mechanisms to tune the collective structure of the colloid dispersion.

\subsection{Transition times of internal states}

\begin{figure}[t]
	\centering
	\includegraphics[width=8.6cm]{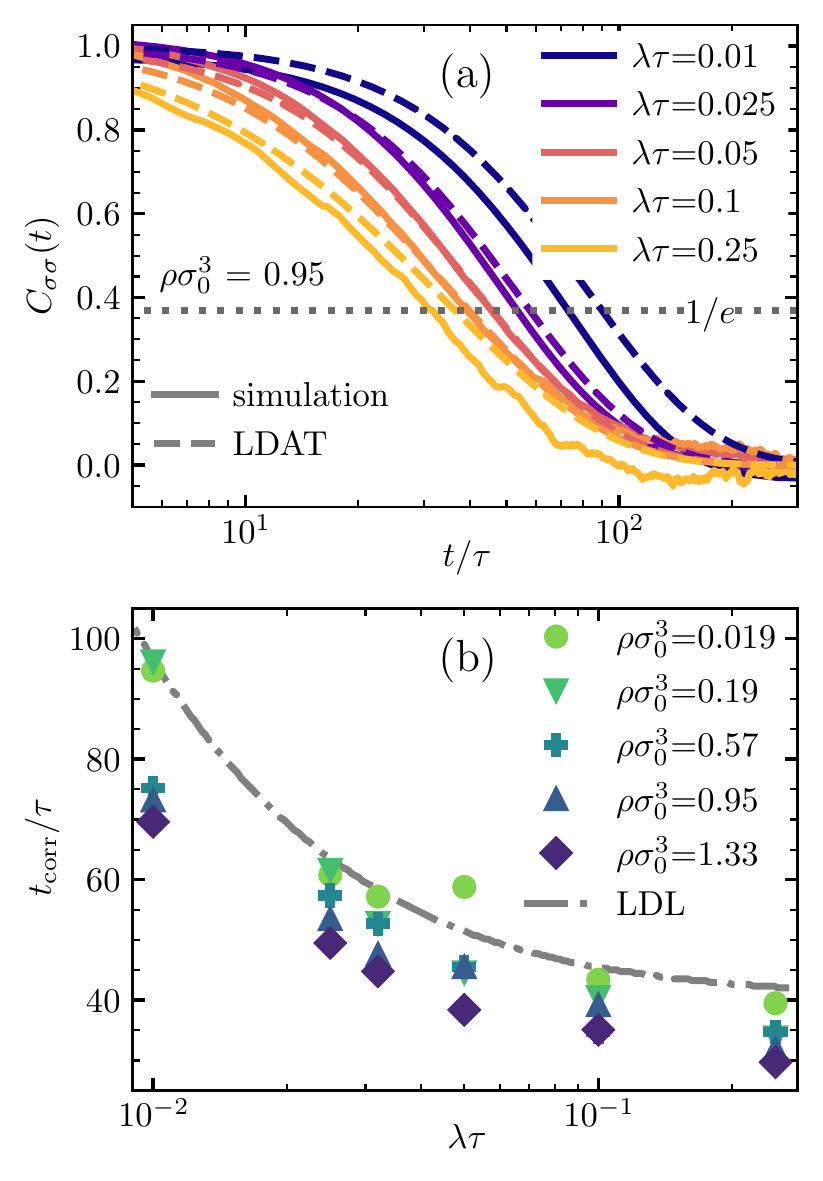}
	\caption{(a) Normalized ACF of the property (see Eq.~(\ref{eq.acf_norm})) for a number density of $\rho\sigma_0^3 = 0.95$ and different switching rates. The solid lines belong to the simulation results, the dashed lines show the theoretical prediction by Eq. (\ref{eq.acf_LDL}) with new potential width $\delta_1$. Intersections with the gray dotted line at $C_{\sigma\sigma} = 1/e$ yield the correlation times. (b) Correlation time $t_\mathrm{corr}$ from Eq.~(\ref{eq.acf_corrTime}) as a function of $\lambda$ for different $\rho\sigma_0^3$. The gray dash-dotted line indicates the theoretical solution in the LDL of Eq.~(\ref{eq.acf_LDL})}
	\label{fig.acf}
\end{figure}

We now turn to the system's property dynamics. For this, we utilize the normalized ACF $C_{\sigma\sigma}(t)$ introduced in Eq.~(\ref{eq.acf_norm}). In Fig.~\ref{fig.acf}(a) it is shown for $\rho\sigma_0^3$ and different switching rates. We can see that the correlation decays faster the lower $\lambda\tau$. This can be explained by the LDL solution in Eq.~(\ref{eq.acf_LDL}): The property relaxes slower for lower switching rates, because drastic changes of the property only appear after a switch. In addition, we can see that the ACF converges for a high $\lambda\tau$ (yellow line) to a simple exponential decay, because the dichotomous time constant $t_\mathrm{D} = 1/2\lambda$ decreases and provides a negligible term in Eq.~(\ref{eq.acf_LDL}). The large fluctuations for high switching rates arise because the initial $\langle \sigma^2 \rangle$ and final value $\langle \sigma \rangle^2$ are very similar for narrow property distributions. To rationalize our findings, we show in Fig.~\ref{fig.acf}(a) also the theoretical ACF calculated from the perturbation theory. It is obtained by inserting the new potential width $\delta_1$ from the LDAT (see Appendix A) into Eq.~(\ref{eq.acf_LDL}). We find a surprisingly good agreement with the simulation which can be traced back to the dominant influence of $\lambda\tau$ in Eq.~(\ref{eq.acf_LDL}) compared to the small changes in $\delta_1$.

The obtained correlation times are shown in Fig.~\ref{fig.acf}(b) as a function of $\lambda\tau$ for different $\rho\sigma_0^3$. The correlation time decreases with increasing $\lambda\tau$; this is true for all covered number densities. In addition, we see that $t_\mathrm{corr}$ decreases also with increasing number density because for larger $\rho\sigma_0^3$ the distribution's width gets narrower (cf. Fig.~\ref{fig.propDist_rho095}). Since the intrinsic time constant, $t_\delta$, in Eq.~(\ref{eq.acf_LDL}) is proportional to the potential's width squared, this leads to an effective larger time constant yielding to a faster decay. Fig.~\ref{fig.acf}(b) also shows as gray dash-dotted line the theoretical $t_\mathrm{corr}$ from the ACF in the LDL (cf. Eq.~(\ref{eq.acf_LDL})). This coincides with the obtained simulation results for low densities (light green circles).

Hence, our results demonstrate how the internal dynamics of colloids are partially controlled by the intrinsic noise. However, to get full control also supervision of the free energy landscape is necessary.

\subsection{Translational diffusion}
\begin{figure}[h]
	\centering
	\includegraphics[width=8.6cm]{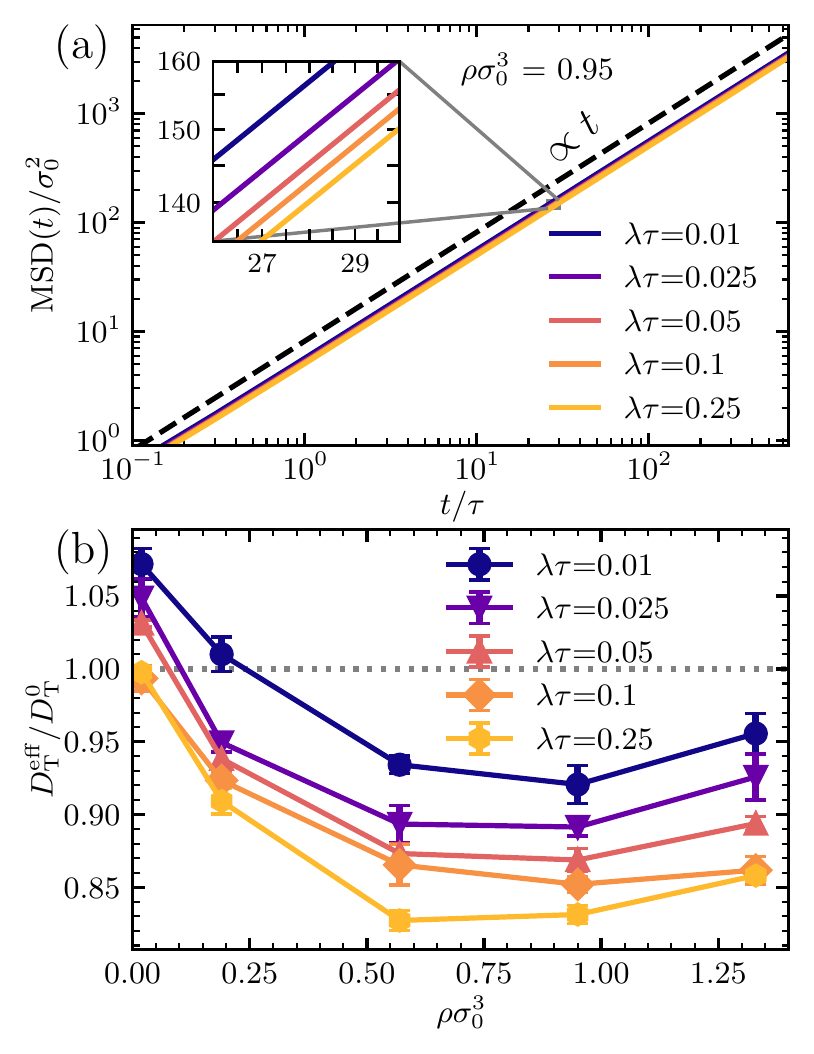}
	\caption{(a) MSD as a function of time according to Eq.~(\ref{eq.msd}) for a high density ($\rho\sigma_0^3 = 0.95$) and different switching rates. The black dashed line shows the slope for a proportionality to time in this log-log plot. (b) Effective diffusion coefficient vs. density for different switching rates. The normalization is done with respect to the diffusion of a free particle with size $\sigma_0$, $D_T^0$.}
	\label{fig.msd}
\end{figure}

Finally, we analyze the translational diffusive dynamics. Fig.~\ref{fig.msd}(a) demonstrates that the MSD in our system is proportional in time and thus normal diffusive. Fig.~\ref{fig.msd}(b) shows the effective diffusion coefficient $D_T^\mathrm{eff}/D_T^0$ vs. $\rho\sigma_0^3$ for different switching rates, where $D_T^0$ denotes the diffusion coefficient of an isolated particle with size $\sigma_0$. The diffusion shows two qualitatively different regimes: For densities up to $\rho\sigma_0^3 \approx 0.9$ the effective diffusion coefficient decreases. This can be explained by crowding which slows diffusion down. The crowding is poorly characterized by the number density because the average particle size shrinks in our model with $\rho\sigma_0^3$. Therefore, crowding is better described by the packing fraction $\eta$ displaying a saturating behavior (cf. Fig.~\ref{fig.propDist_rho095}(c)). As a consequence, for high densities the shrinking particle sizes lead to a higher Stokes  diffusion. This also explains the differences for various switching rates: In the LDL the effective diffusion coefficient is proportional to $\langle 1/\sigma \rangle$ which increases for more bimodal distributions (lower $\alpha$). This is in contrast to existing simulation results of a bidisperse model with switching between a small and a large size \cite{Bley2021_switching2}. The crucial difference is that changes in the switching rate in \cite{Bley2021_switching2} do not change the size distribution.

In summary, the effective diffusion coefficient is comparatively constant under changes in density if one considers that two completely different physical effects (increased crowding and decreased Stokes friction) work against each other. For a specific pair potential it might be even possible to achieve a density independent effective diffusion coefficient. This could be an interesting mechanism for transport of colloids to keep them mobile at higher densities. The results in this section demonstrate that it is possible to control the translational diffusive dynamics by tuning collectively the internal switching rate.

\subsection{Effective temperature definition}
\begin{table}[t]
	\centering
	\vspace{1cm}
	\renewcommand*{\arraystretch}{1.4}
	\begin{tabular}{>{\centering}p{2.5cm}>{\centering}p{1.5cm}<{\centering}p{1.9cm}}
		\toprule
		limit & $T_\mathrm{D}$ & $T_\sigma$ \\ \hline
		$\vD, \lambda \rightarrow \infty$, $\lambda/\vD^2 = \gS/(2T)$ & $\rightarrow T$ & $\rightarrow T$ \\
		$\lambda \rightarrow \infty$ & $\rightarrow 0$ & $\rightarrow 0$ \\
		$\lambda \rightarrow 0$ & $\rightarrow \infty$ & $\rightarrow T(\Delta/\delta)^2$ \\
		\toprule
	\end{tabular}
	\caption{Limits of the definitions of the dichotomous temperature $T_\mathrm{D}$ (cf. Eq.~(\ref{eq.dichotomous_Temperature})) and the property temperature $T_\sigma$ (cf. Eq.~(\ref{eq.property_Temperature})). The analyzed limits are the white-noise limit, as well as the limits for high and low switching rates.}
	\label{tab.res_temp}
\end{table}

In this section, we attempt to define an effective temperature of the property to facilitate an interpretation of the system on a thermodynamic level. This is inspired by already existing approaches to effective temperature definitions in nonequilibrium, e.g., for active motile Brownian particles \cite{Bechinger2016} and two-temperature-baths systems~\cite{Grosberg2015, Dotsenko2013, Netz2020, Polina}. In our system the translational movement has an intrinsic temperature $T$ as input parameter fed into the Gaussian noise; this is not the case for the dichotomous noise. Choosing a standard calculation via the property's mean-squared velocity is not possible since velocities are not considered in an overdamped system. However, the options of defining such a temperature are nonetheless manifold and several of them are discussed in the recent work by Medeiros and Queir\'os \cite{Medeiros2021}.  For simplicity, we consider here an isolated particle to evade the interaction term. Other definitions also include interactions but only for white noise \cite{Grosberg2015}.

For our system we predefine two requirements for a proper temperature: First, in the Gaussian white-noise limit ($\lambda, \vD \rightarrow \infty$ with $\lambda/\vD^2 = \gS/(2T)$) the defined temperature has to converge to the translational temperature $T$, since this describes the thermodynamic equilibrium case. Second, in the no-noise limit ($\lambda \rightarrow \infty$ and $\vD = const.$), representing a frozen property DOF, the temperature has to be zero. In the following we will present and discuss two possible temperature definitions.

The MSD of a free particle in one dimension driven only by dichotomous noise is known \cite{Romanczuk2012} and for long times and finite switching rates linear in time. With the resulting diffusion coefficient $D_\mathrm{D} = \vD^2/2\lambda$ and the Stokes-Einstein equation $D = \kB T/\gamma$ it is possible to define a dichotomous temperature
\begin{equation}
	T_\mathrm{D} = \frac{\gS \vD^2}{2\lambda\kB} = T \frac{(\Delta/\delta)^2}{2\alpha + 2}.
	\label{eq.dichotomous_Temperature}
\end{equation}
This ansatz via the effective diffusion coefficient can also be applied to active Brownian particles \cite{Bechinger2016}. The limits of Eq.~(\ref{eq.dichotomous_Temperature}) are shown in Tab.~\ref{tab.res_temp} and fulfill both our requirements, the Gaussian white-noise limit and the no-noise limit. The name \textit{dichotomous temperature} refers to the fact that this temperature definition is only noise-dependent and does not include the free energy landscape.

Another possible definition was introduced in \cite{Medeiros2021} by comparing definitions from kinetic theory, system entropy and response theory. In the overdamped case they consistently result in the same temperature, for us the \textit{property temperature} \footnote{The temperature definition of Eq.~(\ref{eq.property_Temperature}) and the ones in \cite{Medeiros2021} differ by a factor of 2. In \cite{Medeiros2021} the kinetic term is not considered because velocities are not defined in the overdamped limit. However, to fulfill our condition in the Gaussian limit, we assume that the system's kinetic and potential energy are identical (virial theorem).}
\begin{equation}
T_\sigma = \frac{\beta\delta^2\vD^2\gS^2/\kB}{1 + 2\beta\delta^2\gS\lambda} = T \frac{(\Delta/\delta)^2}{2\alpha + 3}.
\label{eq.property_Temperature}
\end{equation}
It indeed defines a proper temperature which fulfills both of our requirements (see Tab.~\ref{tab.res_temp}). It is worth mentioning that this temperature definition obeys  the fluctuation-dissipation relation between response function $R(t, t')$ and unnormalized auto-correlation function $\partial \tilde{C}_{\sigma\sigma}(t, t')/\partial t' = T R(t, t')$ which holds in thermal equilibrium \cite{Medeiros2021}.

Both definitions predict a qualitative trend for the distributions which is known for temperatures: For low temperatures the probability distribution in a confined potential is narrow (cf. $\alpha=9.0$ in Fig.~\ref{fig.dm}(b)). With increasing temperature the probability distribution gets broader. In the limit of no switching ($\lambda \rightarrow 0$, $\vD = const.$) the interpretations of the two temperatures are more difficult. The term `noise' is also questionable because it is a constant term. While the dichotomous temperature $T_\mathrm{D}$ diverges, the property temperature converges to $T (\Delta/\delta)^2$. Nonetheless, the property $\sigma$ in our system does not even change in time in this limit, so a vanishing temperature could also be justified. However, the permanent energy supply of the internal noise to maintain the size feels conflictive with a cold temperature. Also, the diffusion never reaches proportionality $\mathrm{MSD} \propto t$ for long times. In a nutshell, is there even a reasonable temperature for the low switching rate limit?

We introduced two effective temperature definitions to describe our additional property degree of freedom, the dichotomous temperature $T_\mathrm{D}$ and the property temperature $T_\sigma$. The definitions are, even though from completely different ansatzes, very similar (see Eqs.~(\ref{eq.dichotomous_Temperature}) and (\ref{eq.property_Temperature})). But which one is better? This question cannot be answered. Both definitions have their justification. The dichotomous temperature characterizes the activity of the noise while the property temperature describes the property motion in the harmonic potential. Both definitions fulfill the requirements of the Gaussian white-noise limit ($T_i \rightarrow T$) and the convergence towards zero in the high switching rate limit. At the same time both definitions yield strange results in the low switching rate limit; both in different ways.

The discussion shows that a reasonable temperature definition of a system with dichotomous noise is difficult and can be controversial. Nonetheless, an effective temperature definition gives us the possibility to compare it to the translational temperature and have an effective parameter to describe qualitative trends.

\section{Conclusion}
In summary, we have introduced and characterized a new model for (non-motile) active responsive colloids (ARCs) by assuming that non-Gaussian dichotomous noise governs the internal fluctuations of the particles. This leads to the dichotomous ARC (D-ARC) model.

The intrinsic noise is controlled by a dichotomous velocity and internal switching rate, leading already for a single-particle with a harmonic confinement of the internal DOF to nontrivial parent distributions and dynamic unimodal-bimodal transitions. We emphasize that these transitions, first appearing of purely mathematical nature, have already interesting consequences when interpreted in a physicochemical context. It enables 'living' colloidal particles to control their own size distribution by internally tuning their switching rate and swelling velocity for a possible adaption of function~\cite{Cesar2017}.  The switching rate is indeed an important parameter for biological systems, like bacteria, to regulate transitions between different phenotypes \cite{Dubnau2006}. In addition, we in particular demonstrate that the modification of intrinsic noise (which an active particle can do), has substantial physical effects on the collective behavior of the dispersion: 

\begin{itemize}

\item The intrinsic noise parameters (swelling/shrinking velocity and switching rate) induce a transition from unimodal to bimodal behavior, significantly controlling single particle behavior.

\item The noise-controlled property distribution and its transition is modified by packing and crowding; in turn, the collective liquid structure and dynamics is affected and tuned and property distributions self-consistently modified.

\item Diffusion is homeostatic in the compressible RCs; it can be actively tuned by the intrinsic noise and is relatively constant over the tested density range.

\end{itemize}
Hence, as a key message of this paper, the type of internal fluctuations can play a substantial role not only for single particles but also for the structure and dynamics of the whole interacting dispersion. The main ingredient of the dichotomous noise leading to the observed behavior is the temporary persistence in the direction of motion. We expect other noises with this characteristic to show qualitatively similar results. Examples in two dimensions are Active Brownian Particles or the parental active model \cite{Caprini2022}. We demonstrated a similarly striking behavior already for a simpler model of ARCs where the internal DOF was coupled to a different temperature bath (while still with white noise) than the translation~\cite{Polina}. In contrast to the latter work, however, introducing a colored noise with more 'control buttons' showed much more complexity because of the richer internal structure -- including a uni- to bimodal transition -- and dynamics of a single particle.

An interesting study for the future could be the coupling of the individual (particle-dependent) noise and the internal DOF, leading to feedback. This could lead to collective oscillations of existing self-oscillating colloids \cite{Narita2013} or other synchronous behavior as they happen in bacteria \cite{Wang2015}. The question therefore automatically arises: how can the 'decision' of a single particle lead to collective changes by chemical signaling? -- another proven concept in bacterial and animal kingdoms.

	\section{Acknowledgments}

\begin{acknowledgments}
	We thank Michael Bley, Polina Gaindrik and Sebastian Milster for useful discussions. J.D. acknowledges support by the state of Baden-Württemberg through bwHPC and the German Research Foundation (DFG) through Grant No. INST 39/963-1 FUGG (bw-ForCluster NEMO) and by the DFG via Grant No. WO 2410/2-1 within the framework of the Research Unit FOR 5099 ``Reducing Complexity of Nonequilibrium Systems".
\end{acknowledgments}


\section{Appendix A: Perturbation theory in the low density approximation}
We attempt to compare the simulation results to a theoretical prediction. Therefore, we calculate observables within a simple low density approximation theory (LDAT), which essentially is a perturbation approach for the free energy starting from the LDL. We start with the pair-property distribution function $g(r; \sigma, \sigma')$ which is the conditional probability for a particle with size $\sigma$ to find a particle with size $\sigma'$ at a distance $r$. By approximating that this pair-property distribution function is given by the LDL expression we obtain~\cite{Lin2020}
\begin{equation}
g(r; \sigma, \sigma') \approx \exp\left[-\beta \phi(r; \sigma, \sigma') \right].
\label{eq.lda}
\end{equation}
This is an equilibrium assumption which we apply to our out-of-equilibrium system. We will see that our system is in a quasi-equilibrium making this a reasonable approach for low densities. To obtain the radial distribution function out of the $\sigma$-resolved $g(r; \sigma, \sigma')$ one has to integrate out the properties $\sigma$ and $\sigma'$ \cite{Lin2020}
\begin{equation}
g(r) = \int \mathrm{d}\sigma \int \mathrm{d}\sigma' N(\sigma) N(\sigma') g(r; \sigma, \sigma'),
\label{eq.lda_gofr}
\end{equation}
where $N(\sigma)$ is the emergent property distribution.

The aim of our perturbation theory is to approximate this emergent distribution, since many observables can be calculated or approximated with it (e.g., RDF, packing fraction, size ACF). The ansatz we choose is identical to the one used in \cite{bimodal}, however, we have to modify it at some point. The mean force on a particle with property $\sigma$ originating from interactions with other particles is given as
\begin{equation}
\label{eq.lda_Fpp}
F_\mathrm{pp}(\sigma) = -\rho \int_V \mathrm{d}^3r \int_{-\infty}^{\infty} \mathrm{d}\sigma' N(\sigma') \frac{\partial \phi(r; \sigma, \sigma')}{\partial \sigma} g(r; \sigma, \sigma').
\end{equation}
This is the force $-\partial\phi/\partial\sigma$ between two particles with sizes $\sigma$, $\sigma'$ and distance $r$ where the variables $r$ and $\sigma'$ are integrated out. As shown in \cite{bimodal}, by applying the LDL (Eq.~(\ref{eq.lda})), Eq.~(\ref{eq.lda_Fpp}) can be written as
\begin{equation}
F_\mathrm{pp}(\sigma) = - \frac{5}{4} \pi\rho\epsilon\kappa\left[\sigma^2 + 2\sigma\langle\sigma\rangle + \langle\sigma^2\rangle\right]
\label{eq.LDA_force_result}
\end{equation}
with $\kappa\approx 6.377\times10^{-4}$ for $\beta\epsilon = 500$ and $\langle\cdot\cdot\rangle$ denoting the ensemble average with respect to the emergent distribution.

The discontinuity of the parent distribution at the boundaries $\sigma_0 \pm \Delta$ makes the further procedure as in \cite{bimodal} unreasonable for our system. Therefore, we choose a numerical approach to obtain the emergent distribution. We assume that the disturbed property distribution can still be described by Eq.~(\ref{eq.dm_propDist_Delta_alpha}) but with a new center $\sigma_1$, a new width $\Delta_1$, a new exponent $\alpha_1$ and within a harmonic potential with new width $\delta_1$. This ansatz requires a quadratic energy landscape. To be self consistent, we therefore linearize Eq.~(\ref{eq.LDA_force_result}) by doing a first order Taylor expansion around $\sigma_1$ of the quadratic term $\sigma^2 \approx 2\sigma_1\sigma - \sigma_1^2$. This results in a quadratic term for the interparticle free energy term $\mathcal{F}_\mathrm{pp}(\sigma) = -\int_{0}^{\sigma} \mathrm{d}\sigma'F_\mathrm{pp}(\sigma')$. By inserting the expectation values of the emergent distribution
\begin{equation}
\langle\sigma\rangle = \sigma_1 \qquad \mathrm{and} \qquad \langle\sigma^2\rangle = \sigma_1^2 + \frac{\Delta_1^2}{2\alpha_1 + 3}
\end{equation}
into Eq.~(\ref{eq.LDA_force_result}), we obtain a function with only two unknowns ($\sigma_1$, $\Delta_1$). The initially free parameters $\delta_1 = \delta/\sqrt{1+5\pi\rho\beta\epsilon\kappa\delta^2\sigma_1}$ and $\alpha_1 = \frac{\lambda \Delta_1}{\vD} - 1$ can be eliminated by the definition of the new quadratic potential and the relations in Eq.~(\ref{eq.dm_propDist_Delta_alpha}), respectively. Together with the linear single-particle force term $F_\mathrm{sp} = -\partial_\sigma U(\sigma)$ we get the total mean force $F_\mathrm{tot} = F_\mathrm{sp} + F_\mathrm{pp}$. The latter has to fulfill again the two boundary conditions
\begin{equation}
F_\mathrm{tot}(\sigma_1 - \Delta_1) = v_\mathrm{D} \quad \mathrm{and} \quad F_\mathrm{tot}(\sigma_1 + \Delta_1) = -v_\mathrm{D},
\label{eq.theory_conditions}
\end{equation}
because sizes outside these boundaries are not possible in the steady state. We can solve the two equations in (\ref{eq.theory_conditions}) numerically for $\sigma_1$ and $\Delta_1$ for different particle densities $\rho$ and therefore obtain a LDAT for $N(\sigma; \rho)$.

The general theoretical approach is quite universal and can easily be adapted to other interaction potentials, noises and single-particle potentials \cite{bimodal}. The dependence of Eq.~(\ref{eq.LDA_force_result}) on the potential strength and the hard-sphere limit are discussed in \cite{bimodal}.


	
	\bibliographystyle{apsrev4-2}
	\bibliography{literature_dichotom}
	
\end{document}